\documentclass[a4,authoryear]{elsarticle}
\usepackage{graphicx}
\usepackage{amsmath}
\usepackage{url}
\usepackage{color}
\begin{document}
\date{\empty}
\title{Patterns of Regional Travel Behavior: An Analysis of Japanese Hotel
Reservation Data}
\author{Aki-Hiro SATO$^{1,2}$ \\
$^1$ {\small Department of Applied Mathematics and Physics, Kyoto University,
 Japan} \\
$^2$ {\small Department of Economics, the University of Kiel, Germany}}
\begin{abstract}
This study considers the availability of room opportunities collected from 
a Japanese hotel booking site. We empirically analyze the daily number 
of room opportunities for four areas. To determine 
the migration trends of travelers, we discuss a finite mixture of Poisson
distributions and the EM-algorithm as its parameter estimation
method. We further propose a method to infer the probability of
opportunities existing for each observation. We characterize
 demand-supply situations by means of relationship between the averaged
 room prices and the probability of opportunity existing.
\end{abstract}

\maketitle
\section{Introduction}
Recent technological development enables us to purchase various kinds of
items and services via E-commerce systems. The emergence of Internet
applications has had an unprecedented impact on our life style to
purchase goods and services. From available data of items and services
at E-commerce platforms, we may expect that utilities of agents in 
socio-economic systems are directly estimated.

Such an impact on travel and tourism, specifically, on hotel room
reservations, is significantly considered~\citep{Law}. According to
\citet{Pilia}, 40 per cent of hotel reservations will be made via
Internet in 2008, up from 33 per cent in 2007 and 29 per cent in 2006. 
Therefore, the coverage of room opportunities via the Internet may be
sufficient to provide statistically significant results, and to
conduct a comprehensive analysis based on the hotel booking data collected
from Internet booking sites.
 
From our personal experience, it is found that it is becoming more
popular to make reservations of hotels via the Internet. When we use a hotel
booking site, we notice that we sometimes find preferable room
opportunities or not. Namely, the hotel accessibility seems to be
random. We further know that both the date and place of stay are
important factors to determine the availability of room opportunities. 
Hence, the room availability depends on the calendars (weekdays, weekends,
and holidays) and regions.

This availability of the hotel rooms may indicate the future
migration trends of travelers. Therefore, it is worth considering
accumulation of comprehensive data of hotel availability in order to
detect inter-migration in countries.

The migration processes have been intensively studied in the context of 
socio-economic dynamics with particular interest for quantitative
research. Weidlich and Haag proposed the Master equation with transition
probabilities depending on both regional-dependent and time-dependent utility
and mobility in order to describe collective tendency of agent decision
in migration chance~\citep{Haag:84}. 

Since the motivation of migration seems to come from both psychological
and physical factors, the understanding of the dynamics of the migration
is expected to provide useful insights on inner states of 
agents and their collective behavior.

In the present article, we discuss a model to capture behavior of
consumers at a hotel booking site and investigate statistics of the
number of available room opportunities from several perspectives.

This article is organized as follows. In Sec. \ref{sec:data}, we give a
brief explanation of data description collected from a Japanese hotel
booking site. In Sec. \ref{sec:outlook}, we show an outlook of collected
data of the room opportunities. In Sec. \ref{sec:model}, we consider a
model to capture room opportunities and derive a finite mixture of Poisson
distributions from binomial processes. In Sec. \ref{sec:estimation}, we
introduce the EM-algorithm to estimate parameters of the mixture of
Poisson distributions. In Sec. \ref{sec:numerical-study}, we computed
parameter estimates for an artificial data set generated from the mixture
of Poisson distributions. In Sec. \ref{sec:empirical-analysis}, we show
results of the empirical analysis on the room opportunities and discuss
relationship between existing probabilities of opportunities and their
rates. Sec. \ref{sec:conclusion} is devoted to conclusions.

\section{Data description}
\label{sec:data}
In this section, we give a brief explanation of a method to collect
data on hotel availability. In this study, we used a Web API (Application
Programing Interface) in order to collect the data from a Japanese hotel
booking site named Jalan~\footnote{The data are provided by Jalan Web
service.}. Jalan is one of the most popular hotel reservation
services which provide a WebAPI in Japan. The API is an interface code
set which is designed for a purpose to simplify development of
application programs.

\begin{figure}[hbt]
\begin{center}
\includegraphics[scale=0.4]{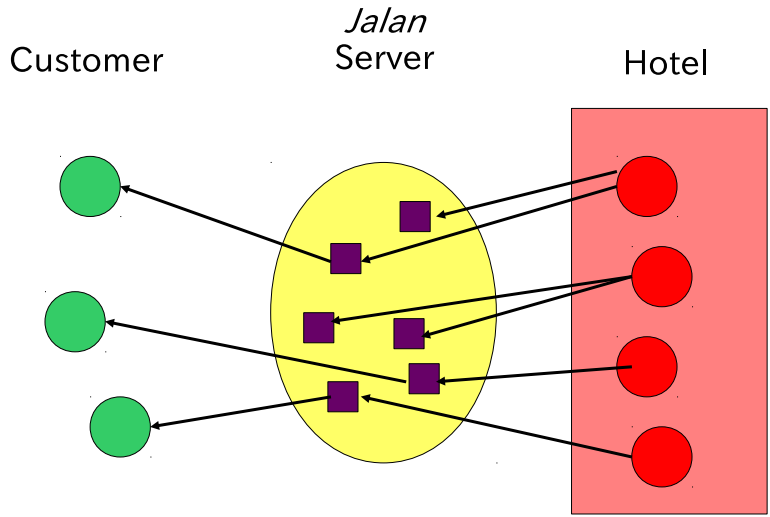}
\end{center}
\caption{A conceptual illustration of Jalan web service. The hotel
managers enter information on rooms (plans) which will be served at the
hotels. Customers can search and book rooms from all the available rooms
(plans) via Jalan web page. }
\label{fig:illustration}
\end{figure}

Jalan Web service provides interfaces for both hotel managers
and customers (see Fig. \ref{fig:illustration}). The mechanism of Jalan
is as follows: The hotel managers can enter information on 
room opportunities served by their hotels via an Internet interface. The
consumers can book rooms from available opportunities via the Jalan Web  
site. The third parties can even built their web services with the Jalan
data by using the Web API.

We are collecting all the available opportunities which appear in Jalan
regarding room opportunities in which two adults will be able 
to stay one night. The data are sampled from the Jalan net web site
({\em http://www.jalan.net}) daily. The data on room opportunities
collected through Jalan Web API are stored as csv files. 

In the data set, there exist over 100,000 room opportunities from over
14,000 hotels. In Tab. \ref{tab:data}, we show contents included in the
data set. Each plan contains sampled date, stay date, regional
sequential number, hotel identification number, hotel name, postal
address, URL of the hotel web page, geographical position, plan name,
and rate.  

Since the data contain regional information, it is possible
for us to analyze regional dependence of hotel rates. Throughout the
investigation, we regard the number of recorded opportunities (plan) as a
proxy variable of the number of available room stocks.

\begin{table}[h]
\caption{The data format of room opportunities.}
\label{tab:data}
\centering
\begin{tabular}{l}
\hline
\hline
Date of collection \\
Date of Stay \\
Hotel identification number \\
Hotel name \\
Hotel name (Kana) \\
Postal code \\
Address \\
URL \\
Latitude \\
Longitude \\
Opportunity name \\
Meal availability \\
the latest best rate \\
Rate per night \\
\hline
\hline
\end{tabular}
\end{table}

\begin{figure}[phbt]
\begin{center}
\includegraphics[scale=0.16]{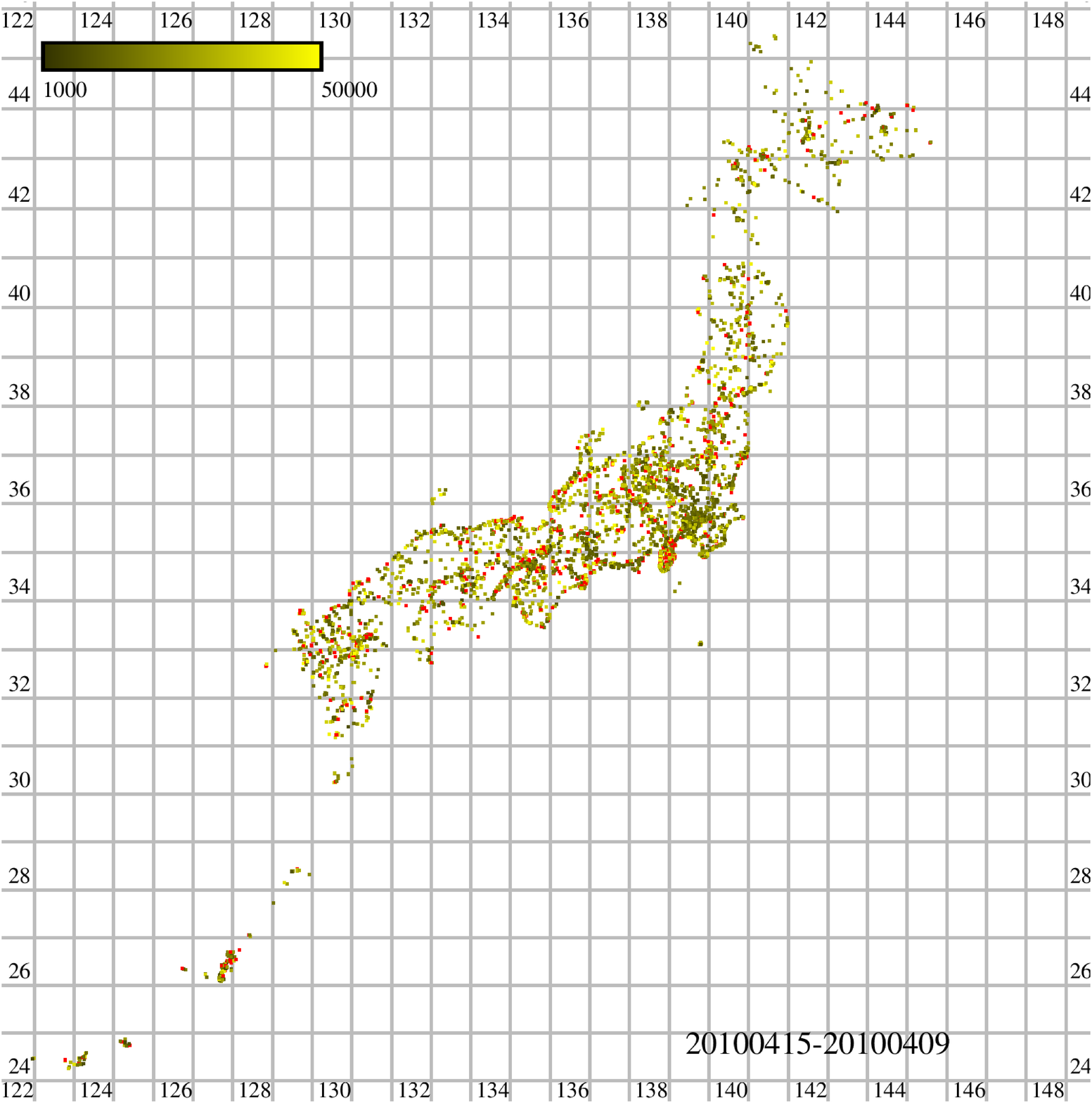}
\includegraphics[scale=0.7]{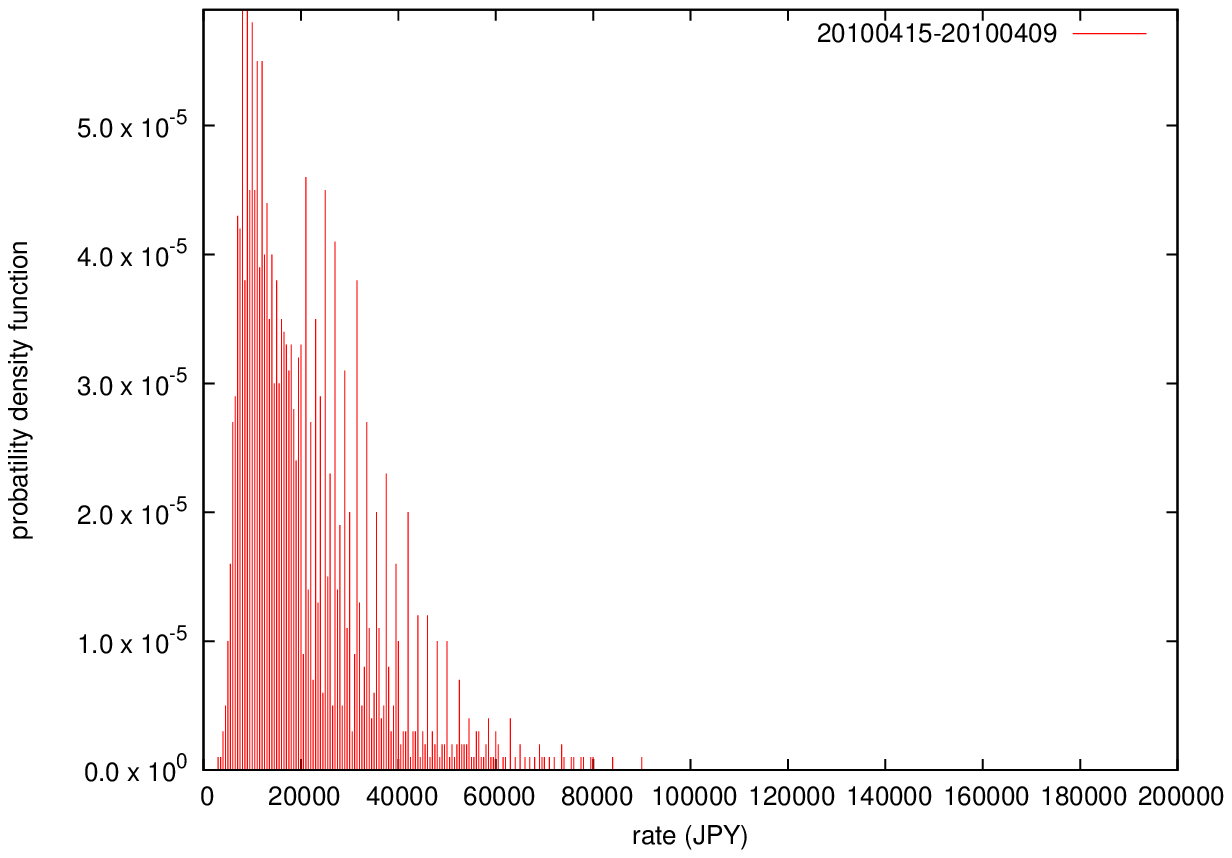}
\end{center}
\caption{An example of rates distributions under the condition that two adults can
stay at the hotel for one night at 15th April 2010 (Top). A
probability density distribution of rates at 15th April 2010 (Bottom).
This data have been sampled on 9th April 2010. Yellow (black) filled
squares represent hotel plans cost 50,000 JPY (1,000) JPY per night. Red
filled squares represent hotel plans cost over 50,000 JPY per night. }
\label{fig:map}
\end{figure}

For this analysis, we used the data for the period from 24th Dec 2009 to
4th November 2010. Fig. \ref{fig:map} shows an example
of distributions and representative rates. The yellow (black) filled squares  
represent hotel plans cost 50,000 JPY (1,000) JPY per night. The red
filled squares represent hotel plans cost over 50,000 JPY per night. We 
found that there is strong dependence of opportunities on
places. Specifically, we find that many hotels are located around
several centralized cities such as Tokyo, Osaka, Nagoya, Fukuoka and so
on. 

Fig. \ref{fig:map} (bottom) shows a probability density distribution on
15th April 2010 all over the Japan. It is found that there are two
peaks around 10,000 JPY and 20,000 JPY on the probability density.

\begin{figure}[hbt]
\begin{center}
\includegraphics[scale=0.9]{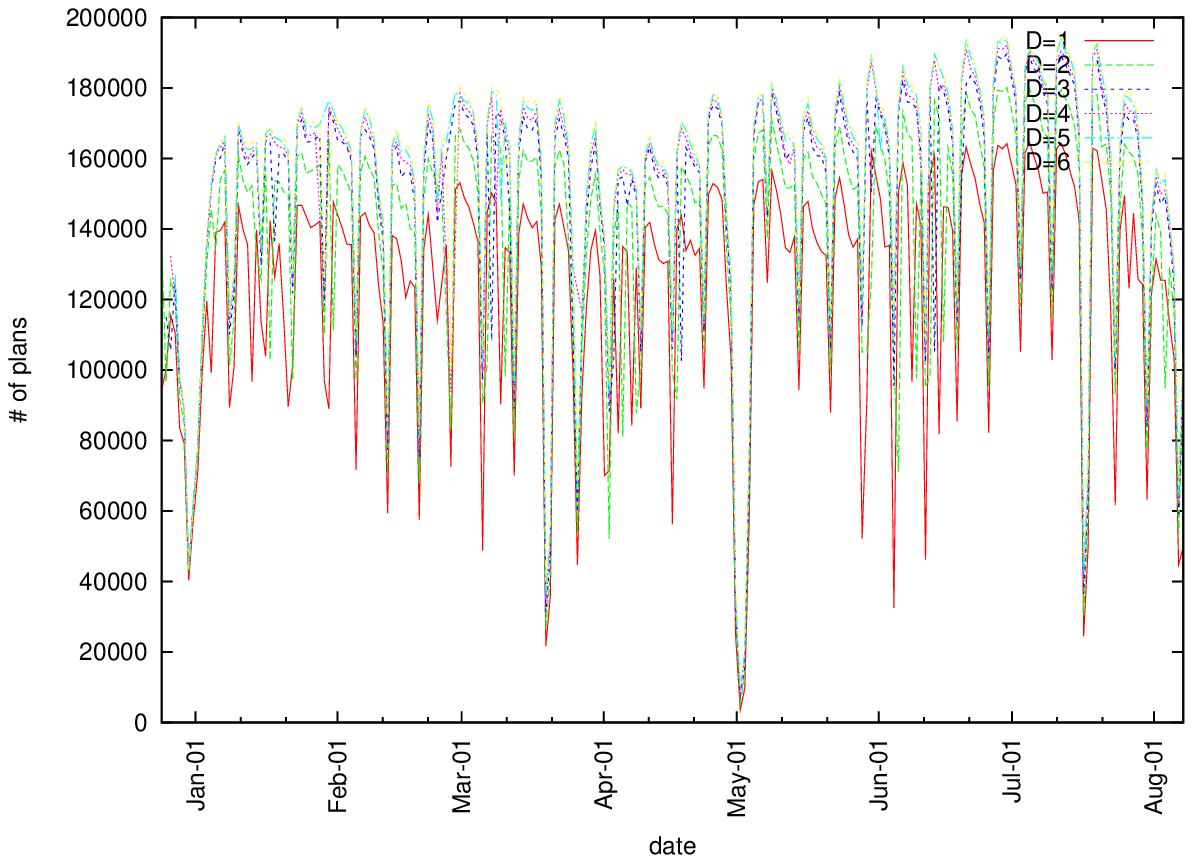}
\end{center}
\caption{The number of hotels in which two adults can stay one night for
 a period from 24th Dec 2009 to 4th November 2010.}
\label{fig:number}
\end{figure}

\section{Overview of the data}
\label{sec:outlook}
The number of room opportunities in which two adults can stay is counted 
from the recorded csv files throughout the whole sampled
period. Fig. \ref{fig:number} shows the daily number of room
opportunities. From this graph, we found three facts:  
\begin{description}
\item{(1)}There exists weekly fluctuation for the number of available room opportunities.
\item{(2)}There is a strong dependence of the number of available
	   opportunities on the Japanese calendar. Namely,
	   Saturdays and holidays drove reservation activities of
	   consumers. For example, during the New Year holidays
	   (around 12/30-1/1) and holidays in the spring season (around
	   3/20), the time series of the numbers show big drops.
\item{(3)}The number eventually increases as the date of stay
	   reaches. Specifically, it is observed that the number of
	   opportunities drastically decreases two days before the date of stay. 
\end{description}
 
\begin{figure}[hbt]
\includegraphics[scale=0.9]{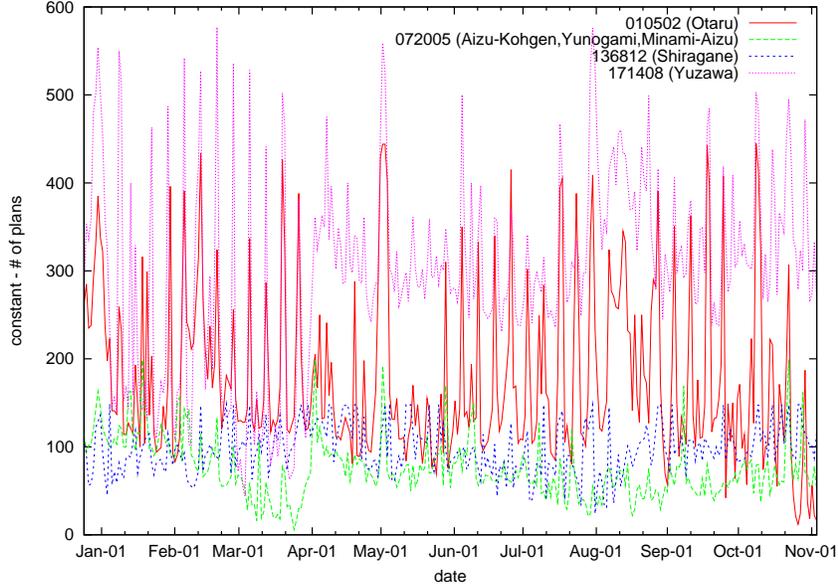}
\caption{The number of demand for four region per
 day. It is found that there exists regional dependence of their
 fluctuations.}
\label{fig:region}
\end{figure}

Fig. \ref{fig:region} shows the number of available room
opportunities at four regions for the period from 24th Dec 2009 to
4th November 2010. We calculated the
numbers at 010502 (Otaru), 072005 (Aizu-Kohgen, Yunogami, and
Minami-Aizu), 136812 (Shiragane), and 171408 (Yuzawa). It is found that
there are regional dependences of their temporal development.

\begin{figure}[hbt]
\centering
\includegraphics[scale=0.8]{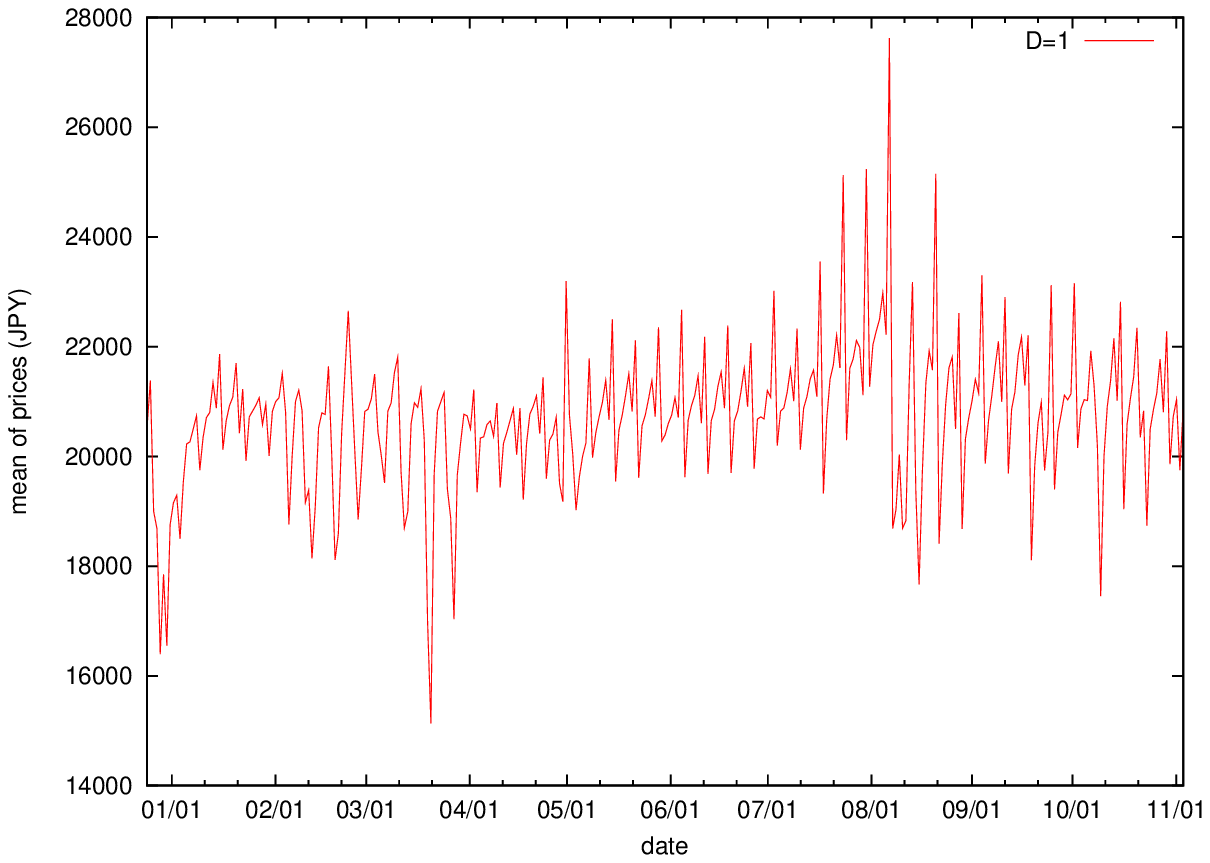}
\caption{Time series of average rates of room
 opportunities on stay dates for four region. The mean value of rates is
 calculated from all the available room opportunities which are observed
 on each stay date.}
\label{fig:mean}
\end{figure}

Furthermore, we show that dependence of averaged rates all over the Japan
on calendar dates in Fig. \ref{fig:mean}. On the New Year holidays in
2010, it is confirmed that the averaged rates rapidly decrease, meanwhile,
on the spring holidays in 2010 the averaged rates rapidly increase. This
difference seems to arise from the difference of consumers motivation
structure and preference on price levels between these holiday seasons. 

\begin{figure}[hbt]
\centering
\includegraphics[scale=0.8]{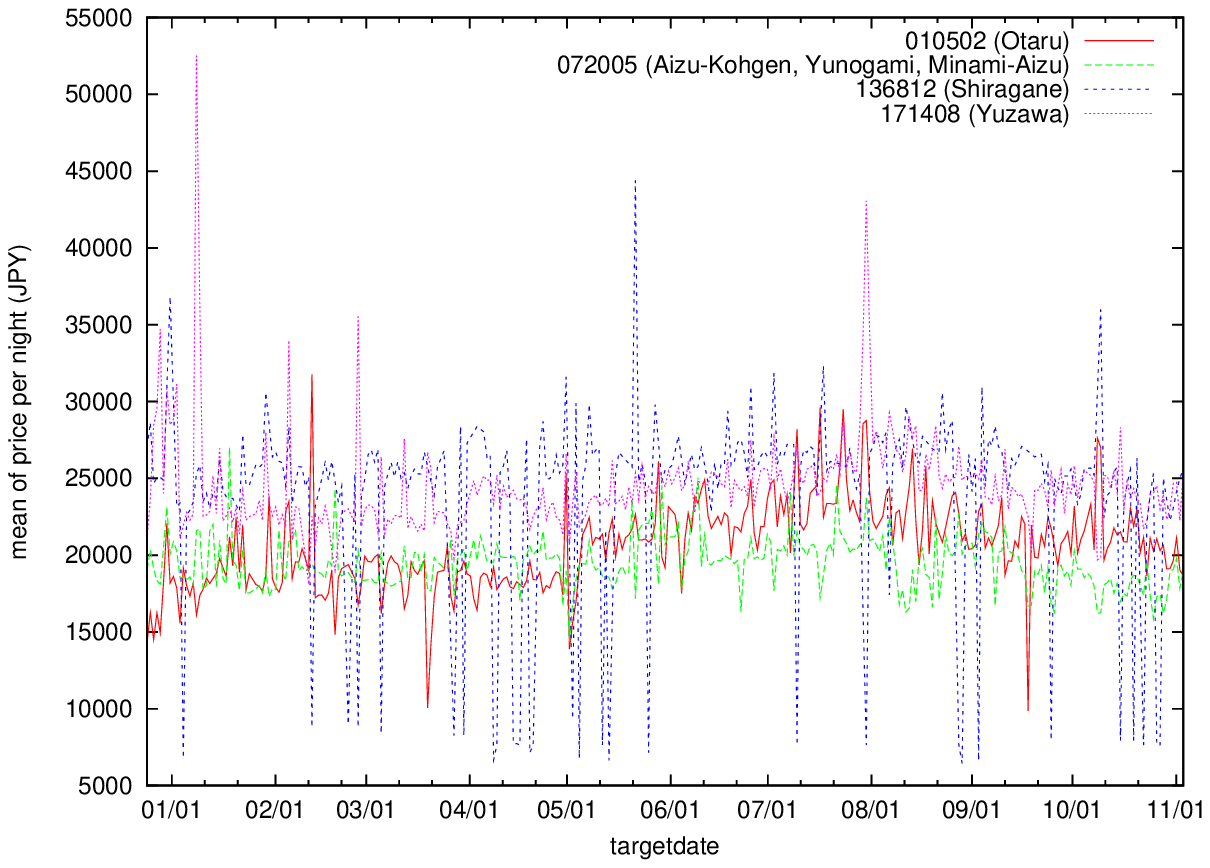}
\caption{Time series of average rates of room
 opportunities on stay dates for four region. The mean value of rates is
 calculated from all the available room opportunities which are observed
 on each stay date.} 
\label{fig:regionalmean}
\end{figure}

Fig. \ref{fig:regionalmean} shows that the dependence of averaged rates
at four regions on calendar dates. The tendency of averaged rates
differs from each other. Specifically, the New Year holidays and the
summer vacation season exhibit such difference. This means that
demand-supply situations depend on regions. We need to know
tendency of the demand-supply situations of each area in a rigorous manner.

\section{Model}
\label{sec:model}
Let $N_m$ and $M$ be the total number of potential rooms at the area $m$
and the total number of potential consumers. The total number of
opportunities $N_m$ may be assumed to be constant since the Internet
booking style has been sufficiently accepted, and almost hotels offer
their rooms via the Internet. Ignoring the birth-death process of
consumers, we also assume that $M$ should be be constant.

We further assume that a Bernoulli random variable
represents booking decision of a consumer from $N_m$ kinds of room
opportunities. In order to express the status of rooms within the
observation period (one day), we introduce $M$ Bernoulli random
variables with time-dependent success probability $p_m(t)$, 
\begin{equation}
y_{mi}(t) =
\left\{
\begin{array}{lll}
1 & w.p. \quad p_m(t) & \mbox{(the $i$-th consumer holds a reservation)}\\
0 & w.p. \quad 1-p_m(t) & \mbox{(the $i$-th consumer does not holds a reservation)}\\
\end{array}
\right., 
\end{equation}
where $y_{mi}(t) \quad (i=1,\ldots,M)$ represents the status of the $i$-th
consumer for a room at time $t$.

If we assume that $p_m(t)$ is sufficiently small, so that $N_m >
\sum_{i=1}^M y_{mi}(t)$, then the number of room opportunities at 
time $t$ may be proportional to the differences between the total number
of potential rooms and the number of booked rooms at time $t$
\begin{equation}
Y_m(t) \propto N_m - \sum_{i=1}^M y_{mi}(t).
\end{equation}
Namely, we have
\begin{equation}
Z_m(t) = kN_m - Y(t) = k\sum_{i=1}^M y_{mi}(t),
\end{equation}
where $k$ is a positive constant.

Assuming further that $y_1(t), \ldots, y_M(t)$ are independently,
identically, distributed, we obtain that the number of available
opportunities $\sum_{i=1}^M y_{mi}(t)$ follows a binomial distribution
$\mbox{B}(M,r_m(t))$. Furthermore, assuming $r_m \ll 1$,
$M \gg 1$, $Mr_m \gg 1$, we can approximate the demand $Z_m(t) = kN_m -
Y_m(t)$ as a Poisson random variable, which follows 
\begin{equation}
\mbox{Pr}_{Z}(l=Z_m|r_m(t)) = \frac{\{Mkp_m(t)\}^l}{l!}e^{-\{Mkp_m(t)\}} = \frac{\{Mr_m(t)\}^l}{l!}e^{-\{Mr_m(t)\}},
\label{eq:Pr-Z}
\end{equation}
where we define $kp_m(t)$ as $r_m(t)$.

Since the agents have some interactions with one another, their
psychological atmosphere (mood), which is collectively created by
agents, influences their decision. Such a psychological effect may be expressed
as probability fluctuations for success probability
$r_m(t)$ at time $t$ in the Bernoulli random variable.
  
Let us assume that the time-dependent probability
$r_m(t) \quad (0 \leq r_m(t) \leq 1)$ is sampled from a
probability density $F_m(r)$. From Eq. (\ref{eq:Pr-Z}),
the marginal distribution for the Poisson distribution conditioning on
$r_m$ with probability fluctuation
$F_m(r_m)$ is given by
\begin{equation}
\mbox{Pr}_{Zm}(l=Z_m) = \int_{0}^{1}F_m(r_m)\frac{(Mr_m)^l}{l!}e^{-Mr_m}dr_m.
\label{eq:dist}
\end{equation}
Since we can observe the number of available opportunities $Z_m(t)$,
we may estimate parameters of the distribution
$F_m(r_m)$ from the successive observations. 

For the sake of simplicity, we further assume that
$r_m(t)$ is sampled from discrete categories $r_{mi}$
with probability $a_{mi}$ ($ 0 \leq r_{mi} \leq 1$; $i =
1, \ldots, K_m; \sum_{i=1}^{K_m} a_{mi} = 1$). These 
parameters are expected to describe motivation structure of consumers
depending on calendar days (weekdays/weekends and special holidays, business
purpose/recreation and so forth). Then since $F_m(r_m)$
is given by
\begin{equation}
F_m(r_m) = \sum_{i=1}^{K_m}a_{mi} \delta (r_m-r_{mi}),
\label{eq:bi}
\end{equation}
$\mbox{Pr}_{Zm}(l=Z_m)$ is calculated as
\begin{eqnarray}
\nonumber
\mbox{Pr}_{Zm}(l=Z_m) &=& \int_{0}^{1}F(r_m)
\frac{(Mr_m)^{l}}{l!}e^{-Mr_m}dr_m, \\ 
       &=& \sum_{i=1}^{K_m} a_{mi} \frac{(Mr_{mi})^{l}}{l!}e^{-Mr_{mi}}.
\label{eq:marginal}
\end{eqnarray}
Hence, Eq. (\ref{eq:marginal}) is concerned with a finite mixture of
Poisson distributions.

\section{Estimation procedure by means of the EM algorithm}
\label{sec:estimation}
The construction of estimators for finite mixtures of distributions has
been considered in the literature of estimation. Estimation procedures
for Poissonian mixture model have been successively studied by several
researchers. Specifically moment estimators and maximum likelihood
estimators are intensively studied. 

The moment estimators were tried on a mixture of two normal distributions by
Karl Peason as early as 1894. Graphical solutions have been given by
\citet{Cassie}, \citet{Harding} and \citet{Bhattacharya}. Rider discusses mixtures of binomial
and mixtures of Poisson distributions in the case of two
distributions~\citep{Rider}.

Hasselblad proposed the maximum likelihood estimator and derived
recursive equations for parameters~\citep{Hasselblad}. The effectiveness
of the maximum likelihood estimator for the mixtures of Poissonian
distributions is widely recognized. Dempster discusses the EM-algorithm
for mixtures of distributions in the several cases~\citep{Dempster}. By 
using the EM-algorithm, we can obtain parameter estimates from mixing data.

Let $z_m(1),\ldots,z_m(T)$ be the number of demand (the
number of potential room opportunities minus the number of available
room opportunities) computed at each observation day. From the
observation sequences, let us consider a method to estimate parameters
of Eq. (\ref{eq:marginal}) based on the maximum likelihood method. In
this case, since the log-likelihood function can be described as
\begin{equation}
L_m(a_{m1},\ldots,a_{mK_m},r_{m1},\ldots,r_{mK_m}) = \sum_{s=1}^{T}\log\Bigl(
\sum_{i=1}^{K_m} a_{mi} \frac{(Mr_{mi})^{z(s)}}{z(s)!}e^{-Mr_{mi}}
\Bigr),
\label{eq:LLF}
\end{equation}
parameter estimates are obtained by the maximization of
the log-likelihood function
$L_m(a_{m1},\ldots,a_{mK_m},r_{m1}\ldots,r_{mK_m})$ 
\begin{equation}
\{\hat{a}_{m1},\ldots,\hat{a}_{mK},\hat{r}_{m1},\ldots,\hat{r}_{mK}\} 
= \underset{\{a_{mi}\},\{r_{mi}\}}{\mbox{arg max}} \quad L_m(a_{m1},\ldots,a_{mK_m},r_{m1}\ldots,r_{mK})
\label{eq:MLE}
\end{equation}
under the constraint $\sum_{i=1}^{K_m}a_{mi}=1$.

The maximum likelihood estimator for the mixture of Poisson models
given by Eq. (\ref{eq:MLE}) can be derived by setting the partial
differentiations of Eq. (\ref{eq:LLF}) with respect to each parameter as
zero (See \ref{sec:derivation-EM}). They lead to the
following recursive equations for parameters;
\begin{eqnarray}
a_{mi}^{(\nu+1)} &=& \frac{1}{T}\sum_{t=1}^T
 \frac{a_{mi}^{(\nu)}F_{mi}^{(\nu)}(z_m(t))}{G_m^{(\nu)}(z_m(t))} 
\quad (i=1,\ldots,K_m), 
\label{eq:a-update}
\\
r_{mi}^{(\nu+1)} &=& \frac{1}{M}\frac{\sum_{t=1}^T
z_m(t) \frac{F_{mi}^{(\nu)}(z_m(t))}{G_m^{(\nu)}(z_m(t))}}{\sum_{t=1}^T
 \frac{F_{mi}^{(\nu)}(z_m(t))}{G_m^{(\nu)}(z_m(t))}} \quad (i=1,\ldots,K_m),
\label{eq:q-update}
\end{eqnarray}
where
\begin{eqnarray}
F_{mi}^{(\nu)}(x) &=& \frac{(Mr_{mi}^{(\nu)})^x e^{-Mr_{mi}^{(\nu)}}}{x!}, \\
G_m^{(\nu)}(x) &=&\sum_{i=1}^{K_m} a_{mi}^{(\nu)} F_{mi}^{(\nu)}(x).
\end{eqnarray}

These recursive equations give us a way to estimate parameters by 
starting from an adequate set of initial values. These recursive 
equations are also referred to as the EM-algorithm for the  mixture of
Poisson distributions~\citep{Dempster,Liu}.

In order to determine the adequate number of parameters, we introduce
the Akaike Information Criteria (AIC), which is defined as
\begin{equation}
AIC(K_m) = 4K_m - 2\hat{L}_m,
\end{equation}
where $\hat{L}_m$ is the maximum value of the log-likelihood function  
in terms of $2K_m$ parameter estimates. $\hat{L}_m$
is computed from the log-likelihood value per observation with parameter
estimates obtained from the EM-algorithm,
\begin{equation}
\hat{L}_m = \sum_{s=1}^{T}\log\Bigl(
\sum_{i=1}^{K_m} \hat{a}_{mi}
\frac{(M\hat{r}_{mi})^{z_m(s)}}{z_m(s)!}e^{-M\hat{r}_{mi}} 
\Bigr).
\end{equation}

Since it is known that the preferred model should be the one with the  
lowest AIC value, we obtain the adequate number of categories $K_m$ as
\begin{equation}
\hat{K_m} = \underset{K_m}{\mbox{arg min}} \quad AIC(K_m).
\end{equation}

Furthermore, we consider the method to determine an underlying Poisson
distribution from which the observation $z_m(s)$ was
sampled. Since the underlying Poisson distribution is one of Poisson
distributions for the mixture, its local likelihood function of
$z_m(s)$ may be maximized over 
all the local likelihood functions of $z_m(s)$. Based
on this idea we propose the following method.

Let $R_{mi}(z) \quad (i=1,\ldots,K_m)$ be
log-likelihood functions of the $i$-th category at area
$m$ with parameter estimate $\hat{r}_{mi}$. From 
Eq. (\ref{eq:marginal}), it is defined as 
\begin{equation}
R_{mi}(z) = z \log M + \log \hat{r}_{mi} - M \hat{r}_{mi}\ \log\Bigl(z!\Bigr).
\end{equation}
By finding the maximum log-likelihood value $R_{mi}(z(s))$
for $i=1,\ldots,K_m$, we can select the adequate
distribution where $z_m(s)$ was extracted. Namely, the
adequate category $\hat{i}_s$ for $z_m(s)$ should be
given as 
\begin{equation}
\hat{i}_s = \underset{i}{\mbox{arg max}} R_i(z_m(s)).
\end{equation}

\section{Numerical simulation}
\label{sec:numerical-study}
Before going into empirical analysis on actual data on room opportunities
with the proposed parameter estimation method, we calculate parameter
estimates for artificial data with it.

We generate the time series $z(s) \quad (s=1,\ldots,T)$ from a mixture of 
Poisson distributions, given by
\begin{equation}
\left\{
\begin{array}{llll}
r(t) &=& r_i \quad w.p. \quad a_i \\
z(t) &\sim& \mbox{Pr}(l=Z(t)|r(t)) = \frac{(Mr(t))^l}{l!}e^{-Mr(t)}
\end{array}
\right..
\end{equation}
where $K$ is the number of categories, $a_i$ represents the probability 
for the $i$-th category to appear $(i=1,\ldots,K; \sum_{i=1}^K a_i=1)$.

\begin{figure}[hbt]
\begin{center}
\includegraphics[scale=0.8]{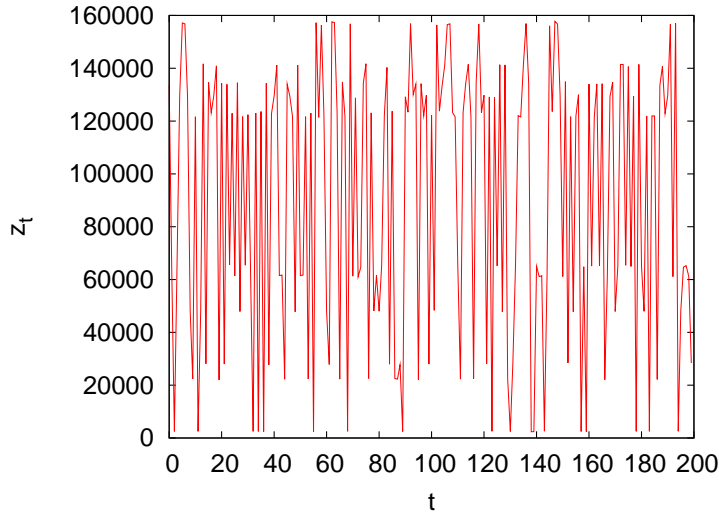}
\end{center}
\caption{Examples of time series generated from the Poissonian mixture
 model for $K=12$ and $M=100,000,000$.}
\label{fig:numerical-ts}
\end{figure}

We set $K=12$ and $M=100,000,000$. Using parameters shown in
Tab. \ref{tab:simulation-parameters}, we generated the artificial data
shown in Fig. \ref{fig:numerical-ts}. Next, we estimated parameters from
$T(=200)$ observations without any prier knowledge on the parameters. 

\begin{figure}[phbt]
\begin{center}
\includegraphics[scale=0.45]{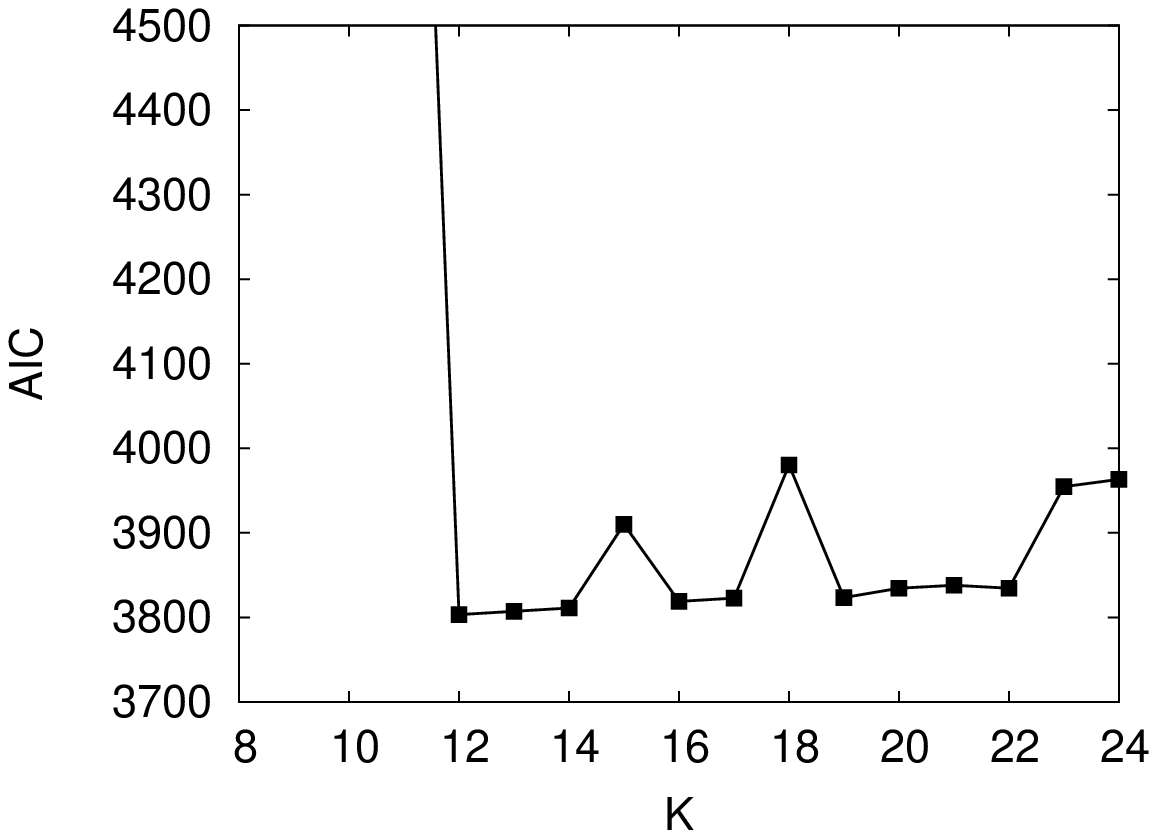}
\includegraphics[scale=0.45]{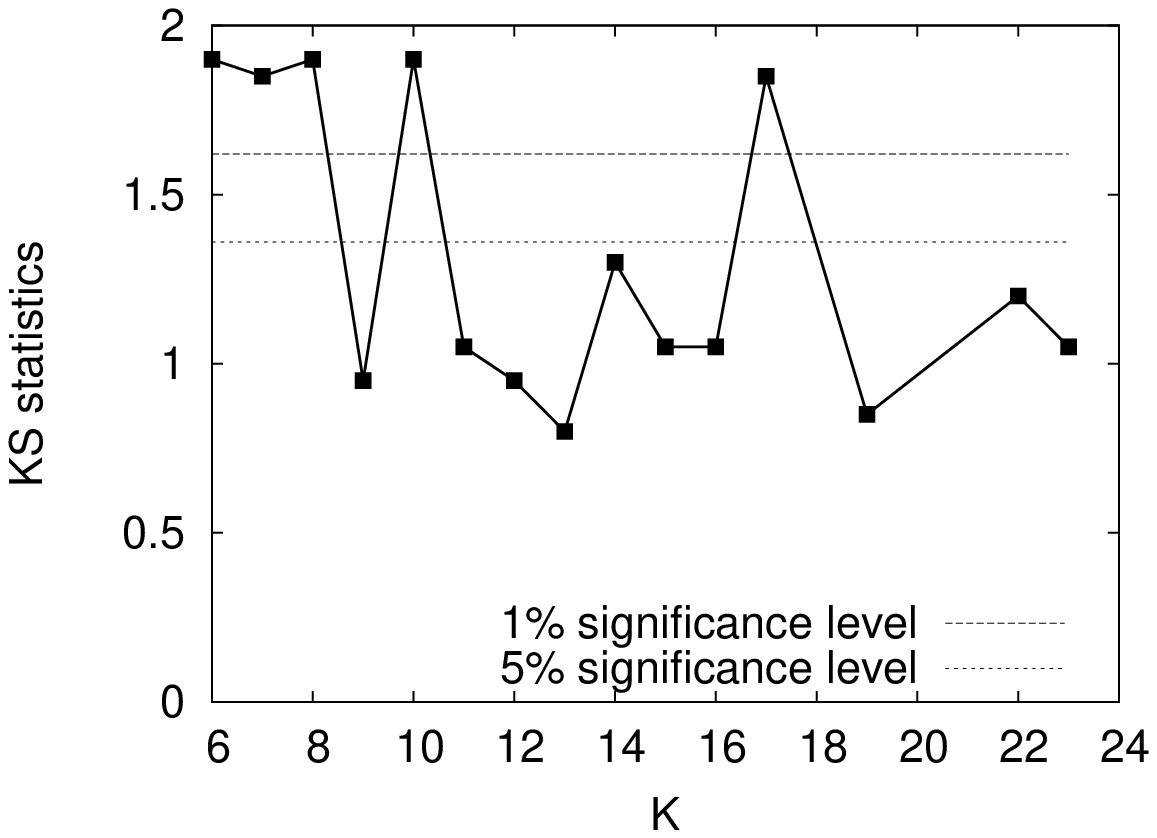}
\end{center}
\caption{The value of AIC for artificial data is shown
 as a function in terms of the number of parameters $K$ (left). The
 lowest value of AIC is found as 3803.20 at $K=12$. The KS statistic
 between the artificial data and estimated ones at each $K$ (right).} 
\label{fig:numerical-AIC}
\end{figure}

As shown in Fig. \ref{fig:numerical-AIC} (top) the AIC values with
respect to $K$ take the minima at $\hat{K}=12$. In order to confirm
adequacy of parameter estimates, we conduct Kolmogorov-Smirnov (KS) test
between the artificial data and sequences of random numbers with
parameter estimates. 

Fig. \ref{fig:numerical-AIC} (bottom) shows KS statistic at each
$K$. Since at $\hat{K}=12$ the KS statistic is computed as 0.327, which
is less than 1.36, the null hypothesis that these time series are
sampled from the same distribution is not rejected at 5\% significance
level. Tab. \ref{tab:simulation} shows parameter estimates.

Furthermore, we selected values of $r_i$ for each observation by means
of the proposed method mentioned in Sec. \ref{sec:estimation}. The
parameter estimates can be computed as a function of time $t$. 

However, we found differences between the parameter estimates and the true
values. Especially, if the close true values of parameters were estimated
as the same parameters. As a result, the number of categories is estimated 
as $\hat{K}=12$, which differs from the number of the true set of
parameters as $K=12$.

\begin{figure}[hbt]
\begin{center}
\includegraphics[scale=0.45]{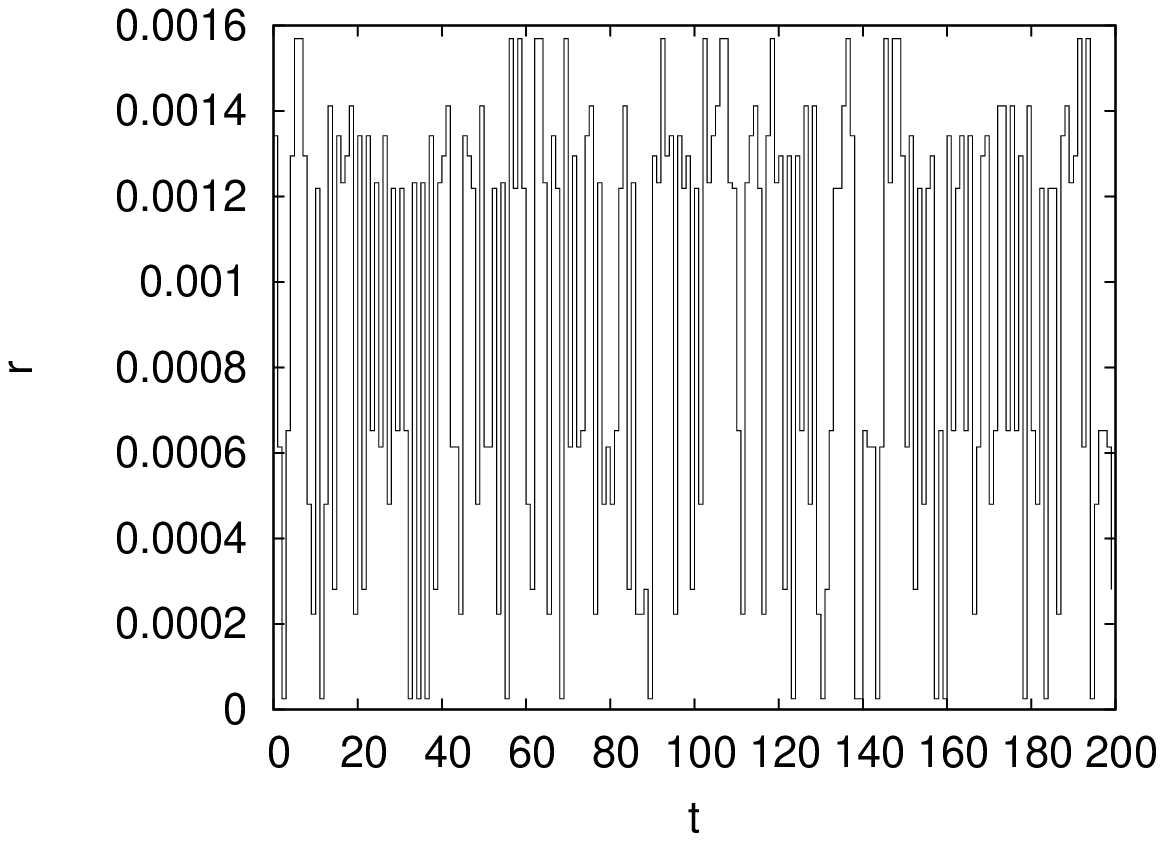}
\includegraphics[scale=0.45]{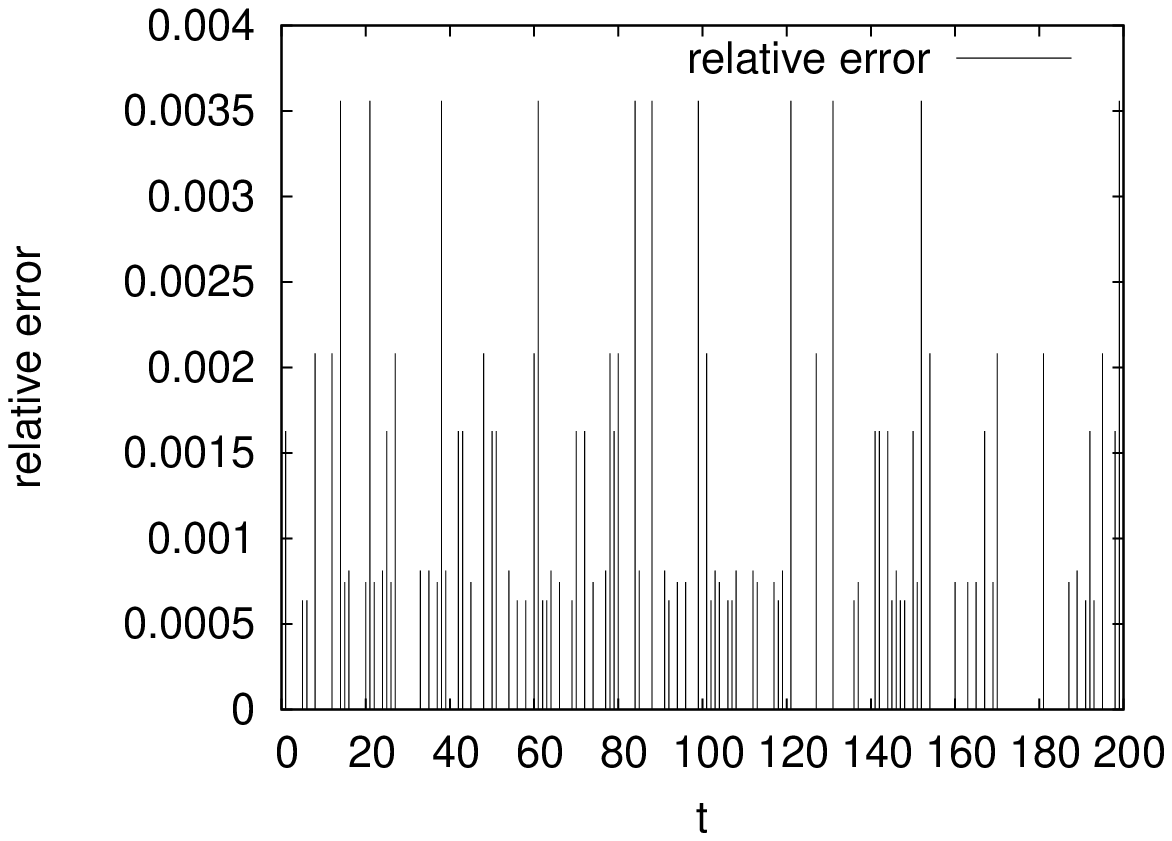}
\end{center}
\caption{Each Parameter estimate for an observation is
 shown as a function in terms of time (left). The relative error between
 the true parameter and the estimated one (right).} 
\label{fig:artificial-estimation}
\end{figure}

After determining the underlying distribution for each observation, we 
further computed estimation errors between the parameter estimates and true
parameters for each observation. As shown in
Fig. \ref{fig:artificial-estimation}, we confirmed that their
estimation error, defined as $|\hat{r}_i(t)-r_i(t)|$ is less than
$8.0\times 10^{-6}$, and that their relative error, defined as
$|\hat{r}_i(t)-r_i(t)|/r_i(t)$, is less than 0.3 \%. It is confirmed that the 
parameter estimates by using the EM-algorithm agree with true values of
parameters for the artificial time series. 

Hence, it is concluded that the discrimination errors between two close
parameters do not play a critical role for the purpose of the parameter
identification at each observation.

\begin{table}[hbt]
\centering
\caption{Parameters of the Poissonian mixture model to generate 
artificial time series. The number of categories is set as $K=12$.}
\label{tab:simulation-parameters}
\centering
\begin{tabular}{lllll}
\hline
\hline
$r_{1}$ & 0.000025 & $a_{1}$ 0.109726 \\
$r_{2}$ & 0.000223 & $a_{2}$ 0.070612 \\
$r_{3}$ & 0.000280 & $a_{3}$ 0.073355 \\
$r_{4}$ & 0.000479 & $a_{4}$ 0.077612 \\
$r_{5}$ & 0.000613 & $a_{5}$ 0.094848 \\
$r_{6}$ & 0.000652 & $a_{6}$ 0.073841 \\
$r_{7}$ & 0.001219 & $a_{7}$ 0.090867 \\
$r_{8}$ & 0.001233 & $a_{8}$ 0.062191 \\
$r_{9}$ & 0.001295 & $a_{9}$ 0.077662 \\
$r_{10}$ & 0.001341 & $a_{10}$ 0.102573 \\
$r_{11}$ & 0.001412 & $a_{11}$ 0.085892 \\
$r_{12}$ & 0.001570 & $a_{12}$ 0.080821 \\
\hline
\hline
\end{tabular}
\end{table}

\begin{table}[hbt]
\centering
\caption{Parameter estimates of the Poissonian mixture model by using the EM
 estimator (bottom). The number of categories was estimated as $\hat{K}=12$ and
 its AIC value is obtained as $AIC = 3803.20$.}
\label{tab:simulation}
\centering
\begin{tabular}{llll}
\hline
\hline
$r_{1}$ & 0.0000247783 $a_{1}$ & 0.0900000000 \\
$r_{2}$ & 0.0002229207 $a_{2}$ & 0.0700000000 \\
$r_{3}$ & 0.0002806173 $a_{3}$ & 0.0550000000 \\
$r_{4}$ & 0.0004798446 $a_{4}$ & 0.0650000000 \\
$r_{5}$ & 0.0006137419 $a_{5}$ & 0.0800000000 \\
$r_{6}$ & 0.0006516237 $a_{6}$ & 0.0950000000 \\
$r_{7}$ & 0.0012185180 $a_{7}$ & 0.1041702140 \\
$r_{8}$ & 0.0012324086 $a_{8}$ & 0.0808297860 \\
$r_{9}$ & 0.0012946718 $a_{9}$ & 0.0850000000 \\
$r_{10}$ & 0.0013420367 $a_{10}$ & 0.1050000000 \\
$r_{11}$ & 0.0014120537 $a_{11}$ & 0.0800000000 \\
$r_{12}$ & 0.0015688622 $a_{12}$ & 0.0900000000 \\
\hline
\hline
\end{tabular}
\end{table}

\section{Empirical results and discussion}
\label{sec:empirical-analysis}
In this section, we apply the proposed method to
estimating parameters for actual data. We estimate the parameters
$a_{mi}$ and $r_{mi}$ from the numbers, which is shown in
Fig. \ref{fig:region} at four regions with the log-likelihood functions
given in Eq. (\ref{eq:LLF}).

Fig. \ref{fig:region} shows estimated time series of the demand
from 24th Dec 2009 to 4th November 2010. In order to obtain this demand, 
we assume that $N_m=\max_{t}\{{Z_m(t)+10}\}$ and that $M$
is approximately equivalent to the total population of Japan, so that $M
= 1,000,000,000$.

\begin{figure}[hbt]
\begin{center}
\includegraphics[scale=0.45]{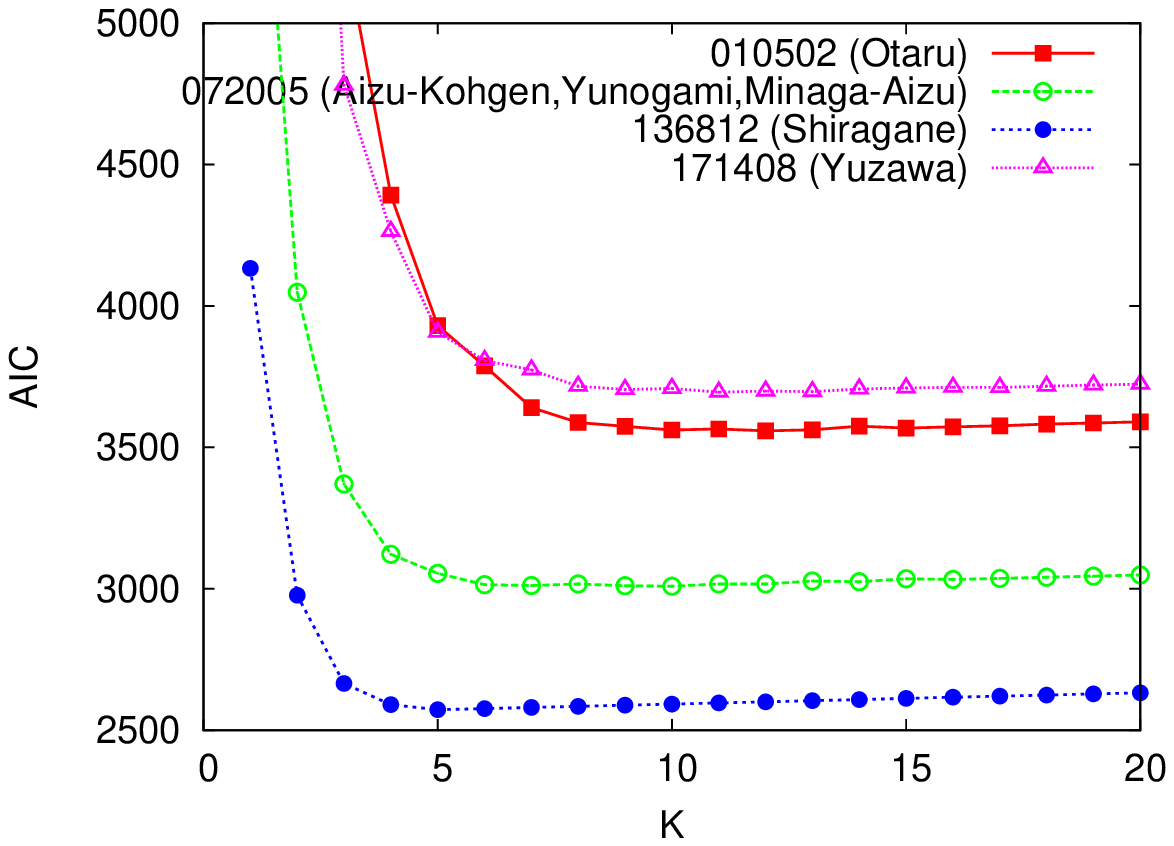}
\includegraphics[scale=0.45]{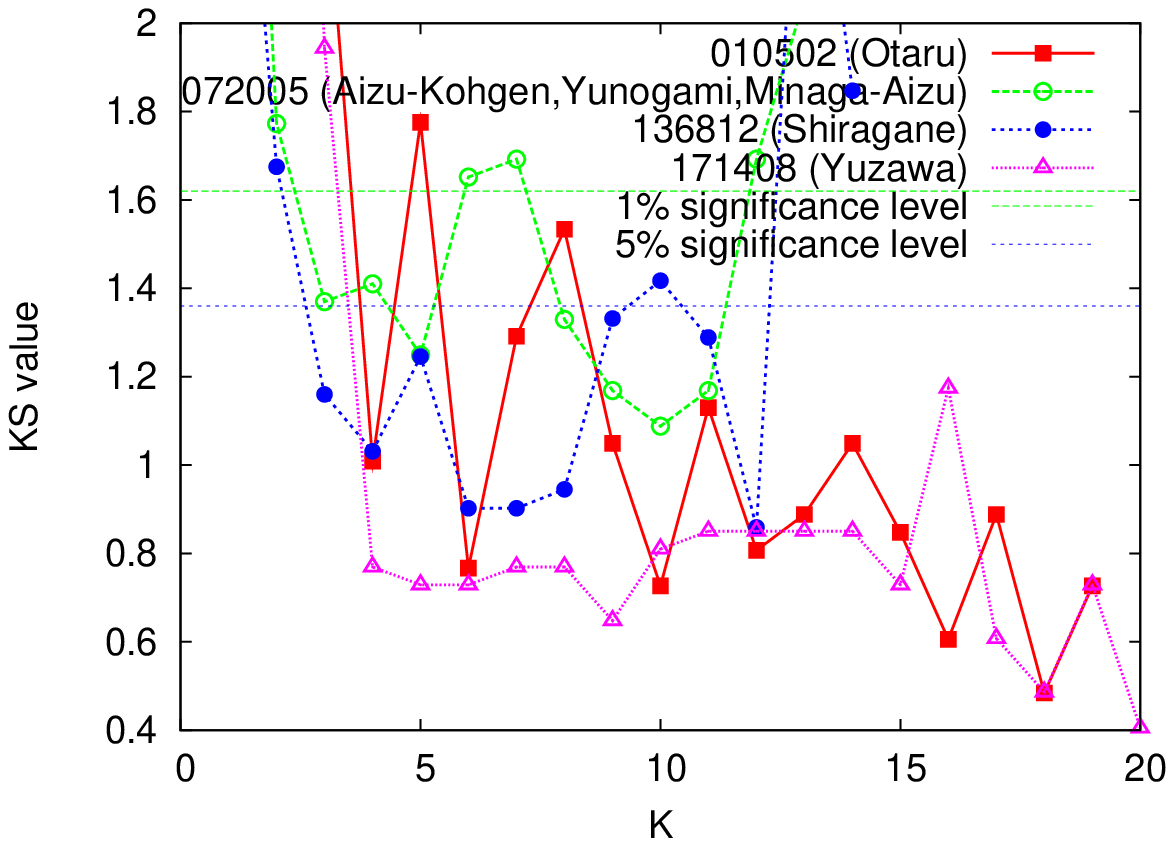}
\end{center}
\caption{The value of AIC (left) and that of KS
 statistics (right) shown as a function in terms of the number of
 parameters $K$ for four areas.} 
\label{fig:AIC-KS}
\end{figure}

According to the values of AIC as shown in Fig. \ref{fig:AIC-KS} (left), the
adequate number of parameters is estimated as $K=12$ (010502), $K=10$
(072005), $K=5$ (136812), and $K=11$ (171408),
respectively. Fig. \ref{fig:AIC-KS} (right) shows the KS value at each
$K$. It is found that the KS test approves the mixture of Poisson
distribution with parameter estimates in statistically
significant. Tab. \ref{tab:AIC-KS} shows the AIC and KS values at the
adequate number of parameters for each area.

\begin{table}[h]
\centering
\caption{At the adequate number of parameters, AIC, maximum log--likelihood
 ($ll$), KS value, and $p$-value for each region.}
\label{tab:AIC-KS}
\begin{tabular}{lllllll}
\hline 
\hline 
regional number & $K$ & $AIC$ & $ll$ & $p$-value & KS value \\
\hline 
010502 & 12 & 3558.31 & 1756.15 & 0.807 & 0.532 \\
072005 & 10 & 3009.10 & 1485.55 & 0.187 & 1.088 \\
136812 & 5 & 2572.33 & 1277.17 & 0.107 & 1.245 \\
171408 & 11 & 3695.25 & 1826.62 & 0.465 & 0.850 \\
\hline 
\hline 
\end{tabular}
\end{table}

\begin{figure}[phbt]
\begin{center}
\includegraphics[scale=0.45]{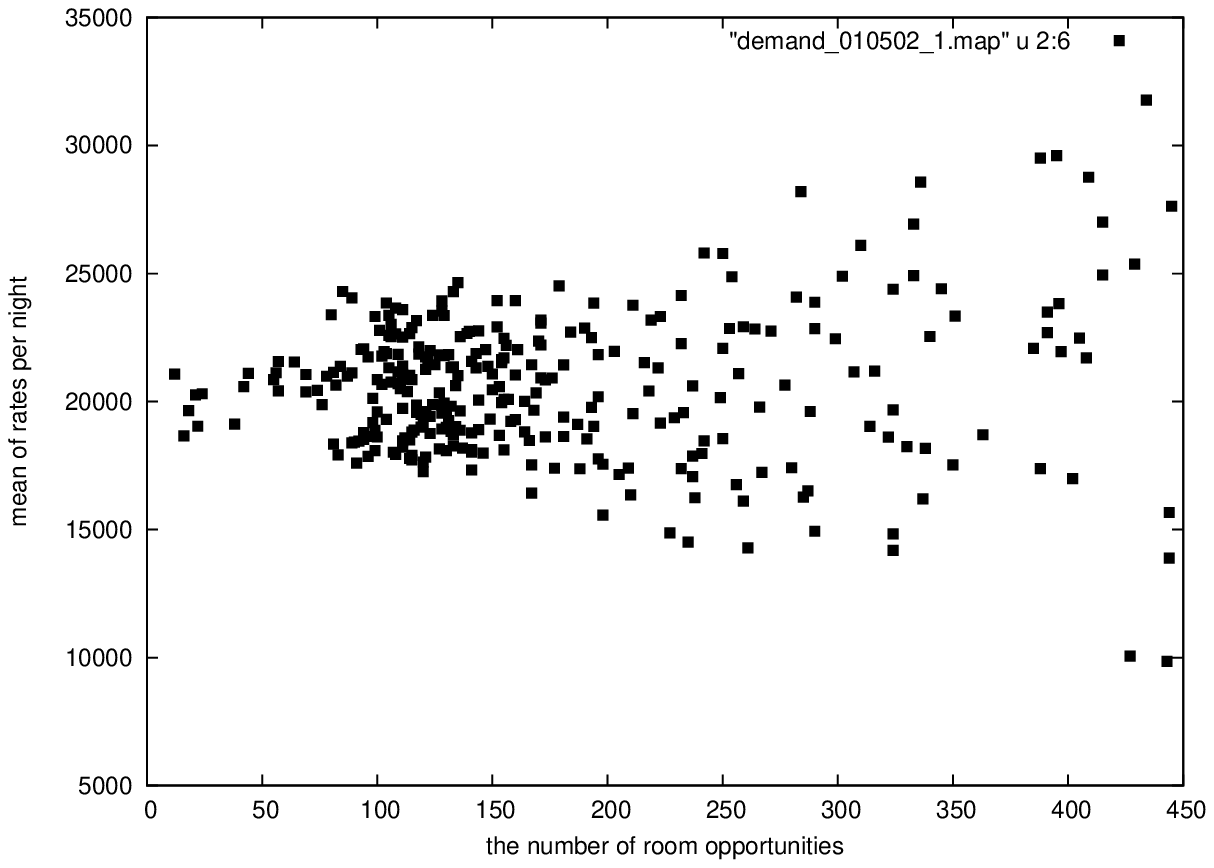}
\includegraphics[scale=0.45]{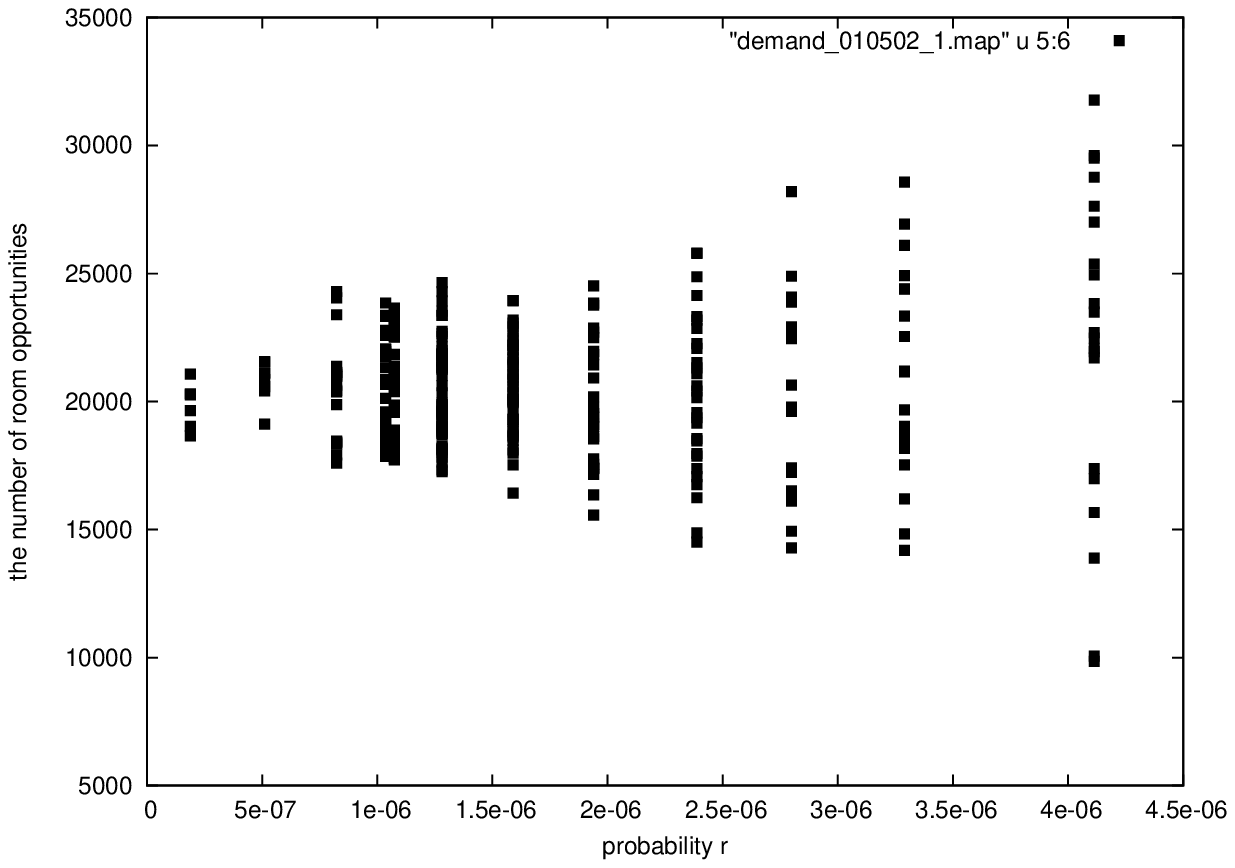}
\end{center}
\caption{The relationship between mean of rates per
 night and the number of opportunities and the relationship between it
 and existence probability for a period from 25th December 2009 to 4th Nov
 2010 (left). Each point represents the relation on each observation day.}
\label{fig:prob-rate}
\end{figure}

Secondly, we confirmed that the relationship between mean of room rates
and the number of opportunities (left) and that between existing
probabilities $r_{mi}$ (right) for each day. Fig. \ref{fig:prob-rate} 
shows their scatter plots during the periods of 25th December 2009 and
4th November 2010. Each point represents their relation for each
day. The variance of room rates proportional to the existence
probability. It is confirmed that the mean of room rates for two adults 
per night is about 20,000 JPY. This means that the excess supply
increases the uncertainty of room rates.

\begin{figure}[hbt]
\includegraphics[scale=0.85]{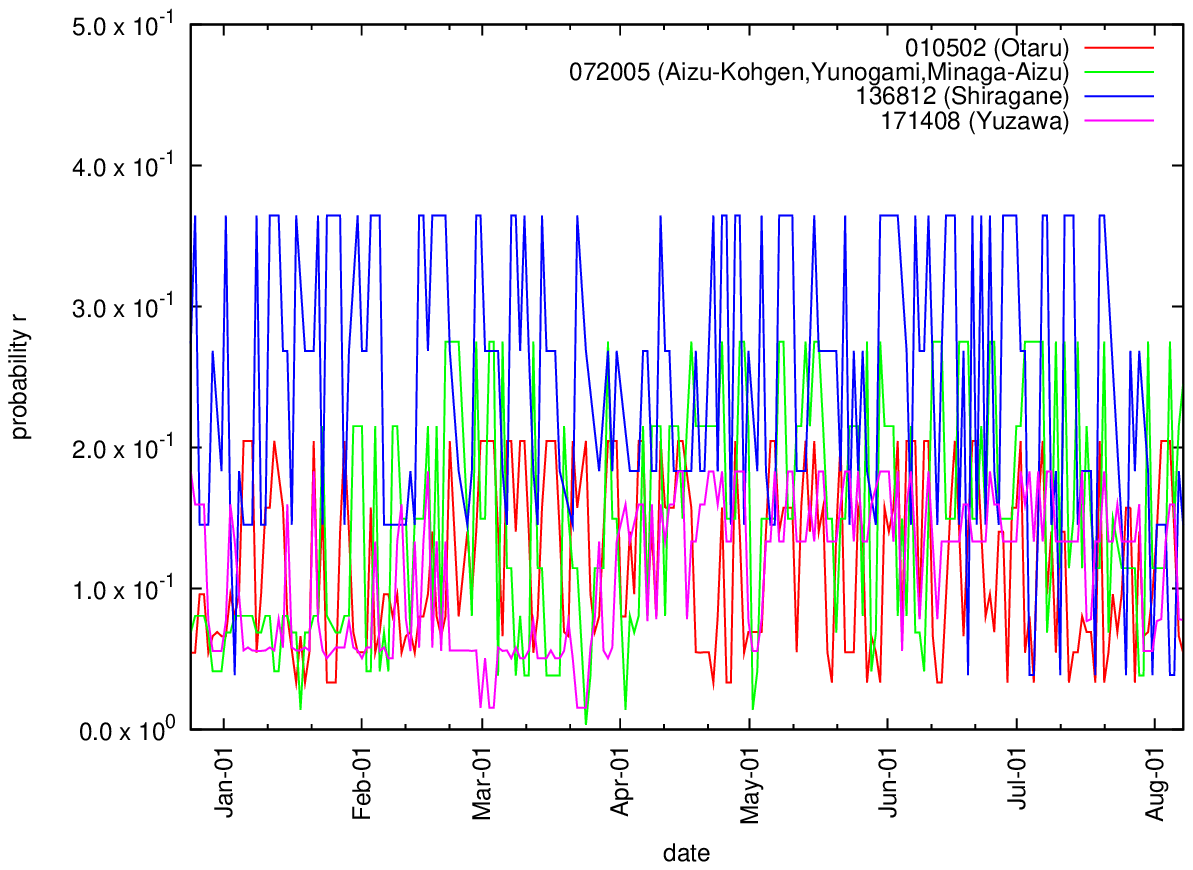}
\caption{The parameter estimates from the number of available
 opportunities on observation date for four regions.}
\label{fig:estimation-region}
\end{figure}

Thirdly, by means of the method to select the underlying distribution,
from Poisson distributions for the mixture, we determined the category
$i$ for each day. As shown in Fig. \ref{fig:estimation-region} (bottom), the
probabilities show strong dependence on the Japanese
calendar.

It is confirmed that there is both regional and 
temporal dependence of the probabilities. We found that the
probabilities take higher values at each region on holidays and weekends
(Saturday). It is observed that higher probabilities maintained in
winter season at 072005 (Aizu-Kohgen,Yunogami,Minami-Aizu). This reason
is because this place is one of winter ski resorts.

Specifically, on holidays and Saturdays, they take smaller
values than on weekdays. Tabs. \ref{tab:empirical-param} and
\ref{tab:empirical} show parameter estimates and exact dates included
in each category at 010502 (Otaru), respectively. From this table we
found that there is travel tendency of this region on seasons. 

It is found that a lot of travelers visited and the hotel rooms
were actively booked at this area on dates included in categories 1
to 3. On the other hand, this area were actively booked on dates
included in categories 10 to 12.

The covariates among the numbers of room opportunities at different 
regions are the important factors to determine the demand-supply
situations all over the Japan. 


From Tab. \ref{tab:empirical-param}, it is confirmed that in the case of
Otaru the end of October to the beginning of November in 2010 was highly
demanded season. This tendency is different from the calendar dates. The
relationship between the number of opportunities and averaged rates is
slightly different from that between the existence probability and the
averaged rates. By using our proposed method we can compare the
difference of comsumers' demand between the dates. From the dependence
of averaged prices on the probability $r_i$, we can understand
preference and motivation structure of consumers for travel and tourism.

\begin{table}[hbt]
\centering
\caption{Parameter estimates of the Poissonian mixture model by using the EM
 estimator. The number of categories were estimated as $\hat{K}=12$ at 010502.}
\label{tab:empirical-param}
\begin{tabular}{clcl}
\hline 
\hline 
$r_{1}$ & 0.0000001884 & $a_{1}$ & 0.0195519998 \\
$r_{2}$ & 0.0000005108 & $a_{2}$ & 0.0245098292 \\
$r_{3}$ & 0.0000008236 & $a_{3}$ & 0.0548239374 \\
$r_{4}$ & 0.0000010361 & $a_{4}$ & 0.0332800534 \\
$r_{5}$ & 0.0000010742 & $a_{5}$ & 0.1572987311 \\
$r_{6}$ & 0.0000012821 & $a_{6}$ & 0.2045505810 \\
$r_{7}$ & 0.0000015900 & $a_{7}$ & 0.1402093740 \\
$r_{8}$ & 0.0000019395 & $a_{8}$ & 0.0801500614 \\
$r_{9}$ & 0.0000023878 & $a_{9}$ & 0.0959105486 \\
$r_{10}$ & 0.0000027989 & $a_{10}$ & 0.0544931969 \\
$r_{11}$ & 0.0000032900 & $a_{11}$ & 0.0661506482 \\
$r_{12}$ & 0.0000041136 & $a_{12}$ & 0.0690710390 \\
\hline
\hline 
\end{tabular}
\end{table}

\begin{table}[p]
\caption{The dates included in each category for 010502 (Otaru).}
\label{tab:empirical}
{\small 
\begin{tabular}{|l|p{12cm}|}
\hline
1 & 2010-10-26,2010-10-27,201-10-28,2010-11-01,2010-11-03,2010-11-04 \\
\hline
2 & 2010-09-01,2010-09-26,2010-09-30,2010-10-05,2010-10-18,2010-10-25,2010-10-31,2010-11-02\\
\hline
3 & 2010-02-01,2010-02-02,2010-02-03,2010-04-19,2010-04-21,2010-04-22,2010-05-12,2010-05-19,2010-05-23,2010-05-24,2010-05-25,2010-05-30,2010-07-14,2010-07-15,2010-07-15,2010-07-22,2010-07-31,2010-08-31,2010-09-06,2010-10-12,2010-10-12,2010-10-29 \\
\hline
4 & 2010-01-11,2010-01-12,2010-01-15,2010-01-24,2010-02-04,2010-03-15,2010-03-23,2010-04-12,2010-04-13,2010-04-14,2010-04-18,2010-04-25,2010-05-09,2010-05-11,2010-05-13,2010-05-18,2010-05-26,2010-06-01,2010-06-03,2010-06-16,2010-06-30,2010-07-01,2010-07-06,2010-07-26,2010-07-27,2010-08-30,2010-09-15,2010-09-16,2010-09-20,2010-09-28,2010-10-04,2010-10-17,2010-10-21 \\
\hline
5 & 2010-01-06,2010-01-07,2010-01-08,2010-01-13,2010-01-22,2010-01-29,2010-02-22,2010-03-01,2010-03-02,2010-03-03,2010-03-04,2010-03-07,2010-03-08,2010-03-10,2010-03-11,2010-03-16,2010-03-17,2010-03-18,2010-03-22,2010-03-25,2010-03-30,2010-03-31,2010-04-01,2010-04-06,2010-04-07,2010-04-11,2010-04-15,2010-04-16,2010-04-28,2010-05-06,2010-05-07,2010-05-14,2010-05-16,2010-05-22,2010-06-04,2010-06-06,2010-06-07,2010-06-08,2010-06-10,2010-06-11,2010-06-17,2010-06-21,2010-06-22,2010-07-02,2010-07-20,2010-08-03,2010-08-04,2010-08-05,2010-08-17,2010-08-20,2010-08-24,2010-09-07,2010-09-08,2010-09-12,2010-09-21,2010-09-22,2010-10-08,2010-10-20\\
\hline
6 & 2010-01-28,2010-01-30,2010-02-18,2010-02-26,2010-02-28,2010-03-05,2010-03-09,2010-03-12,2010-03-19,2010-03-29,2010-04-04,2010-04-09,2010-04-29,2010-05-05,2010-05-08,2010-05-15,2010-05-17,2010-05-21,2010-05-27,2010-06-02,2010-06-18,2010-06-20,2010-06-23,2010-06-27,2010-06-28,2010-07-09,2010-07-11,2010-07-12,2010-07-29,2010-08-02,2010-08-06,2010-08-19,2010-08-23,2010-08-29,2010-09-05,2010-09-17,2010-09-23,2010-09-27,2010-09-29,2010-10-01,2010-10-02,2010-10-19\\
\hline
7 & 2010-01-04,2010-01-16,2010-01-23,2010-02-09,2010-02-15,2010-02-16,2010-02-19,2010-02-21,2010-02-24,2010-03-14,2010-03-28,2010-04-02,2010-04-03,2010-04-10,2010-04-24,2010-06-09,2010-06-24,2010-07-04,2010-07-16,2010-08-22,2010-09-03,2010-09-10,2010-09-14,2010-09-24,2010-10-22,2010-10-30\\
\hline
8 & 2009-12-24,2010-12-27,2010-12-28,2010-01-03,2010-01-05,2010-01-10,2010-02-07,2010-02-08,2010-02-10,2010-02-17,2010-02-27,2010-03-26,2010-04-05,2010-04-08,2010-06-25,2010-07-08,2010-07-23,2010-07-25,2010-08-01,2010-08-11,2010-08-15,2010-08-16,2010-08-18,2010-08-21,2010-08-25,2010-10-07,2010-10-15,2010-10-16\\
\hline
9 & 2009-12-25,2009-12-26,2009-12-29,2010-01-09,2010-01-21,2010-02-05,2010-02-11,2010-02-14,2010-03-13,2010-04-20,2010-04-30,2010-07-03,2010-07-10,2010-08-08,2010-08-09,2010-08-10,2010-08-12,2010-08-26,2010-08-27 \\
\hline
10 & 2009-12-30,2010-01-01,2010-01-02,2010-01-19,2010-02-12,2010-02-20,2010-03-06,2010-03-21,2010-05-29,2010-06-05,2010-06-12,2010-06-19,2010-07-30,2010-08-07,2010-08-13,2010-08-14,2010-09-04,2010-09-11,2010-10-11,2010-10-23 \\
\hline
11 & 2009-12-31,2010-02-06,2010-02-13,2010-03-20,2010-03-27,2010-05-01,2010-05-02,2010-05-03,2010-05-04,2010-06-26,2010-07-17,2010-07-18,2010-07-24,2010-07-31,2010-08-28,2010-09-18,2010-09-19,2010-09-25,2010-10-09,2010-10-10 \\
\hline
12 & 2010-01-18,2010-01-20,2010-01-25,2010-01-27,2010-04-23,2010-04-26,2010-04-27,2010-05-20,2010-05-28,2010-05-31,2010-06-13,2010-06-14,2010-06-29,2010-07-05,2010-07-13,2010-07-19,2010-07-21,2010-07-28,2010-09-02,2010-09-09,2010-09-13,2010-10-03,2010-10-13,2010-10-14,2010-10-24 \\
\hline
\end{tabular}
}
\end{table}


\section{Conclusion}
\label{sec:conclusion}
We analyzed the data of room opportunities collected from a Japanese
hotel booking site. We found that there is strong dependence of the
number of available opportunities on the Japanese calendar. 

Firstly, We proposed a model of hotel booking activities based on a
mixture of Poisson distributions with time-dependent intensity. From a
binomial model with a time-dependent success probability, we derived
the mixture of Poisson distributions. Based on the mixture model, we
characterized the number of room opportunities at each day with
different parameters regarding their difference as motivation structure
of consumers dependent on the Japanese calendar.

Secondarily, we proposed a parameter estimation method on the basis of the
EM-algorithm and a method to select the underlying distribution for each
observation from Poisson distributions for the mixture through the
maximization of the local log-likelihood value. 

Thirdly, we computed parameters for artificial time series generated
from the mixture of Poisson distributions with the proposed method, and
confirmed that the parameter estimates agree with the true values of
parameters in statistically significant. We conducted an empirical
analysis on the room opportunity data. We confirmed that the
relationship between the averaged prices and the probabilities of
opportunities existing is associated with demand-supply
situations. Furthermore, we extracted multiple time series of the
numbers at four regions and found that the migration trends of
travelers seem to depend on regions.

It was found that these large-scaled data on hotel opportunities enable us
to see several invisible properties of travelers' behavior in Japan. 

As future work, we need to use more high-resolution data on booking of
consumers at each hotel to capture demand-supply situations. If we can
use such data, then we will be able to control room rates based
on consumers' preference. A future emerging technology will make it
possible to see or foresee something which we can not see at this moment.

\section*{Acknowledgement}
This study was financially supported by the Excellent Young Researcher
Overseas Visiting Program (\# 21-5341) of the Japan Society for the
Promotion of Science (JSPS). The author is thankful very much to
Prof. Dr. Thomas Lux for his fruitful discussions and kind
suggestions. The author expresses to Prof. Dr. Dirk
Helbing his sincere gratitude for stimulative discussions. This is a
research study, which has been started in collaboration with
Prof. Dr. Dirk Helbing.

\appendix
\section{Derivation of the EM-algorithm}
\label{sec:derivation-EM}
In this section, we mention a derivation of the EM algorithm from the
maximum likelihood estimation procedure. Let $G_m(z)$ represent a mixture
of $K_m$ Poisson distributions $F_{mi}(z)$
\begin{eqnarray}
F_{mi}(z) &=& \frac{(Mr_{mi})^z}{z!} e^{-Mr_{mi}}, (i=1,\ldots,K_m)
\label{eq:Fi}
\\
G_m(z) &=&\sum_{i=1}^{K_m} a_{mi} F_{mi}(z),
\label{eq:G}
\end{eqnarray}
where $a_{mi}$ denote mixing ratios, which are normalized as
\begin{equation}
\sum_{i=1}^{K_m} a_{mi} = 1.
\label{eq:normalization}
\end{equation}
From observations $\{z_m(t)\}$ the log-likelihood function in terms of the
parameters $a_{mi}, r_{mi} \quad (i=1,\ldots,K_m)$ can be written as
\begin{equation}
L_m(a_{m1},\ldots,a_{mr},r_{m1},\ldots,r_{mK_m}) = \sum_{t=1}^T \log G_m\bigl(z_m(t)\bigr).
\label{eq:ll-mixture}
\end{equation}
Inserting Eqs. (\ref{eq:Fi}) and (\ref{eq:normalization}) into
Eq. (\ref{eq:ll-mixture}), we have
\begin{eqnarray}
\nonumber
L_m(a_{m1},\ldots,a_{mK_m},r_{m1},\ldots,r_{mK_m}) = \\
\sum_{t=1}^T \log \Bigl(
\sum_{i=1}^{K_m-1}a_i \frac{(Mr_{mi})^{z_m(t)}}{z_m(t)!} e^{-Mr_{mi}} +
(1-\sum_{i=1}^{K_m-1}a_{mi}) \frac{(Mr_{mK_m})^{z_m(t)}}{z_m(t)!}
e^{-Mr_{mK}}\Bigr).
\label{eq:ll-mixture2}
\end{eqnarray}
Partially differentiating Eq. (\ref{eq:ll-mixture2}) in terms of $a_{mi}$
and $r_{mi}$, we obtain
\begin{eqnarray}
\frac{\partial L_m}{\partial a_{mi}} &=&
 \sum_{t=1}^T\frac{F_{mi}(z(t))-F_{mK}(z_m(t))}{G_m(z_m(t))} \quad
 (i=1,\ldots,K_m-1),
\label{eq:condition-ai}
\\
\frac{\partial L_m}{\partial r_{mi}} &=&
 \sum_{t=1}^T\frac{a_{mi}F_{mi}(z_m(t))}{G_m(z_m(t))}\Bigl(\frac{z_m(t)}{r_{mi}}-M\Bigr)
 \quad (i=1,\ldots,K_m-1), 
\label{eq:condition-ri}
\\
\frac{\partial L_m}{\partial r_{mK_m}} &=&  
\sum_{t=1}^T\frac{(1-\sum_{i=1}^{K_m}a_{mi})F_K(z_m(t))}{G_m(z_m(t))}\Bigl(\frac{z_m(t)}{r_{mK}}-M\Bigr).
\label{eq:condition-rK}
\end{eqnarray}
Multiplying $a_{mi}$ by Eq. (\ref{eq:condition-ai}) and summing them over
$i$ we have
\begin{equation}
\sum_{t=1}^T \frac{F_{mi}(z_m(t))}{G_m(z_m(t))} = T \quad (i=1,\ldots,K_m),
\label{eq:condition-ai2}
\end{equation}
and multiplying $a_{mi}/T$ by Eq. (\ref{eq:condition-ai2}) we obtain
\begin{equation}
a_{mi} = \frac{1}{T}\sum_{t=1}^T \frac{a_{mi}F_{mi}(z_m(t))}{G_m(z_m(t))} \quad (i=1,\ldots,K_m).
\end{equation}
From Eqs. (\ref{eq:condition-ri}) and (\ref{eq:condition-rK}) we
immediately obtain
\begin{equation}
r_{mi} = \frac{1}{M}\frac{\sum_{t=1}^T
 z_m(t)\frac{F_{mi}(z_m(t))}{G_m(z_m(t))}}{\sum_{t=1}^T \frac{F_{mi}(z_m(t))}{G_m(z_m(t))}} \quad (i=1,\ldots,K_m).
\end{equation}

Therefore, if we find an adequate set of initial values for parameters
$\{a_{mi}^{\{0\}}\}$ and $\{r_{mi}^{\{0\}}\}$, then we can calculate parameters
by using the update rule for $\{a_{mi}^{(\nu)}\}$ and $\{r_{mi}^{(\nu)}\}$
\begin{eqnarray}
a_{mi}^{(\nu+1)} &=& \frac{1}{T}\sum_{t=1}^T 
 \frac{a_{mi}^{({\nu})}F_{mi}^{(\nu)}(z_m(t))}{G_m^{(\nu)}(z_m(t))} \quad (i=1,\ldots,K_m), 
\label{eq:a-update-app}
\\
r_{mi}^{(\nu+1)} &=& \frac{1}{M}\frac{\sum_{t=1}^T
z_m(t) \frac{F_{mi}^{(\nu)}(z_m(t))}{G_m^{(\nu)}(z_m(t))}}{\sum_{t=1}^T
 \frac{F_{mi}^{(\nu)}(z_m(t))}{G_m^{(\nu)}(z_m(t))}} \quad (i=1,\ldots,K_m),
\label{eq:q-update-app}
\end{eqnarray}
where
\begin{eqnarray}
F_{mi}^{(\nu)}(z) &=& \frac{(Mr_{mi}^{(\nu)})^z}{z!} e^{-Mr_{mi}^{(\nu)}}, \\
G_m^{(\nu)}(z) &=&\sum_{i=1}^{K_m} a_{mi}^{(\nu)} F_{mi}^{(\nu)}(z).
\end{eqnarray}

We compute these recursive equations by setting an arbitrary set of
parameters. Some of them are convergent and the others
divergent. Therefore, it is important for us to find an adequate set of
initial values when we use the EM-algorithm given by
Eqs. (\ref{eq:a-update-app}) and (\ref{eq:q-update-app}) 
for estimation. To do so, we use a way to find a candidate of parameters
in a stochastic manner. This procedure consists of three parts. 

The choice of initial values is based on Finch et al.'s
algorithm~\cite{Finch}. Their idea is that, given the mixing proportion
$r_{mi}$, the $s$-th order statistics of observations $z_m(t_s) \quad
(s=1,\ldots,T)$ are separated into $K_m$ parts. Each sub-block contains
the $i$-th $[a_{im}T]$ observations assumed to belong to the $i$-th
component of the mixture. We compute mean of the $i$-th sub-block
$\mu_{mi}$ and use it as initial value $r_{im}^{0}=\mu_{mi}/M$.

In the Monte Carlo step, we randomly allocate $r_{m1}, \ldots, r_{mK_m}$,
where $\sum_{i=1}^{K_m} r_{mi}=1$ and evaluate the log-likelihood function
at the point. If the value of log-likelihood function at this point is
finite, we choose this set of parameters as the new starting point for
the recursive equation.

Further setting the set of parameters as an initial condition, we
recursively calculate the EM-algorithm until the log-likelihood value
converges. If it is greater than the maximum value of log-likelihood
function which has already obtained in the Monte Carlo step,
then the set of parameters as a candidate of parameter estimates.

Repeating this procedure until we can not find any points which improve
the value of log-likelihood function in the Monte Carlo step, we estimate an
adequate set of parameters. This algorithm is described as follows.

\begin{description}
\item{(0)} Set $maxobj = 0$ and $counter=0$.
\item{(1)} Generate normalized random numbers $a'_{mi}$ as $a'_{mi} =
	   b'_{mi}/\sqrt{\sum_{i=1}^{K_m}}b'_{mi}$ by using {\it i.i.d.}
	   uniform random numbers $b'_{mi}$. 
\item{(2)} $r_{mi}$ are generated with Finch et al.'s algorithm from
	   $a'_{im}$. If $counter > MAXCOUNT$, then go to Step (6).
\item{(3)} If $L_m(a'_{m1},\ldots,a'_{mK_m},r'_{m1},\ldots,r'_{mK})$ is
	   greater than $maxobj$, then we set $maxobj$ as the value,
	   $r_{mi}:=r'_{mi}$ and $a_{mi}:=a'_{mi}$, and go to Step
	   (4). Otherwise go to Step (1). 
\item{(4)} From the starting point
	   $(a_{m1}\ldots,a_{mK},r_{m1},\ldots,r_{mK})$,  
	   compute Eqs. (\ref{eq:a-update}) and (\ref{eq:q-update})
	   recursively until the value of log-likelihood function
	   converges.
\item{(5)} If the maximum value of log-likelihood function in terms of 
	   the converged set of parameters is larger than $maxobj$, 
	   then set the value as $maxobj$ and record the
	   solution as a candidate of parameter estimates.
\item{(6)} $counter=counter+1$ and if $counter < MAXCOUNT$, then go
 to (1). Otherwise go to Step (7).
\item{(7)} Stop this computer program and display $maxobj$ and the
	   recorded candidate as parameter estimates.
\end{description}

\end{document}